\documentclass[fleqn,usenatbib]{mnras}

\usepackage{newtxtext,newtxmath}
\usepackage[T1]{fontenc}
\usepackage{ae,aecompl}

\usepackage{graphicx}	% Including figure files
\usepackage{amsmath}	% Advanced maths commands
\usepackage{amssymb}	% Extra maths symbols

\title[Electron-impact excitation of Fe$^+$]{Electron-impact fine-structure excitation of \ion{Fe}{ii} at low temperature}

\author[Yier Wan et al.]{
Yier Wan,$^{1}$
C. Favreau,$^{2}$ 
S. D. Loch,$^{2}$ 
B. M. McLaughlin,$^{3}$ 
Yueying Qi$^{4}$
and P. C. Stancil$^{1}$\thanks{E-mail: stancil@physast.uga.edu}
\\
$^{1}$Department of Physics and Astronomy, Center for Simulational Physics,
The University of Georgia, Athens, GA 30602, USA\\
$^{2}$Department of Physics, Auburn University, Auburn, AL 36849, USA\\
$^{3}$Centre for Theoretical Atomic and Molecular Physics (CTAMOP), 
School of Mathematics and Physics, Queen's University Belfast,
Belfast BT7 1NN, UK\\
$^{4}$College of Mathematics, Physics and Information,
Jiaxing University, Jiaxing, Zhejiang, 314001, PRC\\
}

\date{Accepted XXX. Received YYY; in original form ZZZ}

\pubyear{2019}

\begin{document}
\label{firstpage}
\pagerange{\pageref{firstpage}--\pageref{lastpage}}
\maketitle

\begin{abstract}
\ion{Fe}{ii} emission lines are observed from nearly all classes of astronomical objects 
over a wide spectral range from the infrared to the ultraviolet. To meaningfully interpret these lines, reliable atomic data are necessary. In work presented here we focused on low-lying fine-structure transitions, within the ground term, due to electron impact.  We provide effective collision strengths together with estimated uncertainties as functions of temperature of astrophysical importance ($10 - 100,000$ K). Due to the importance of fine-structure transitions within the ground term, the focus of this work is on obtaining accurate rate coefficients at the lower end of this temperature range, for applications in low temperature environments such as the interstellar medium. We performed three different flavours of scattering calculations: i) a intermediate coupling frame transformation (ICFT) $R$-matrix method, ii) a Breit-Pauli (BP) $R$-matrix method, and iii) a Dirac $R$-matrix method. The ICFT and BP $R$-matrix calculations involved three different AUTOSTRUCTURE target models each. The Dirac $R$-matrix calculation was based on a reliable 20 configuration, 6069 level atomic structure model. Good agreement was found with our BP and Dirac $R$-matrix collision results compared to previous $R$-matrix calculations. We present a set of recommended effective collision strengths for the low-lying forbidden transitions together with associated uncertainty estimates.

\end{abstract}


%%%%%%%%%%%%%%%%%%%%%%%%%%%%%%%%%%%%%%%%%%%%%%%%%%

%%%%%%%%%%%%%%%%% BODY OF PAPER %%%%%%%%%%%%%%%%%%%
\section{Introduction}
\label{sec:introduction}

\begin{figure}
\centering
\includegraphics[width=\linewidth]{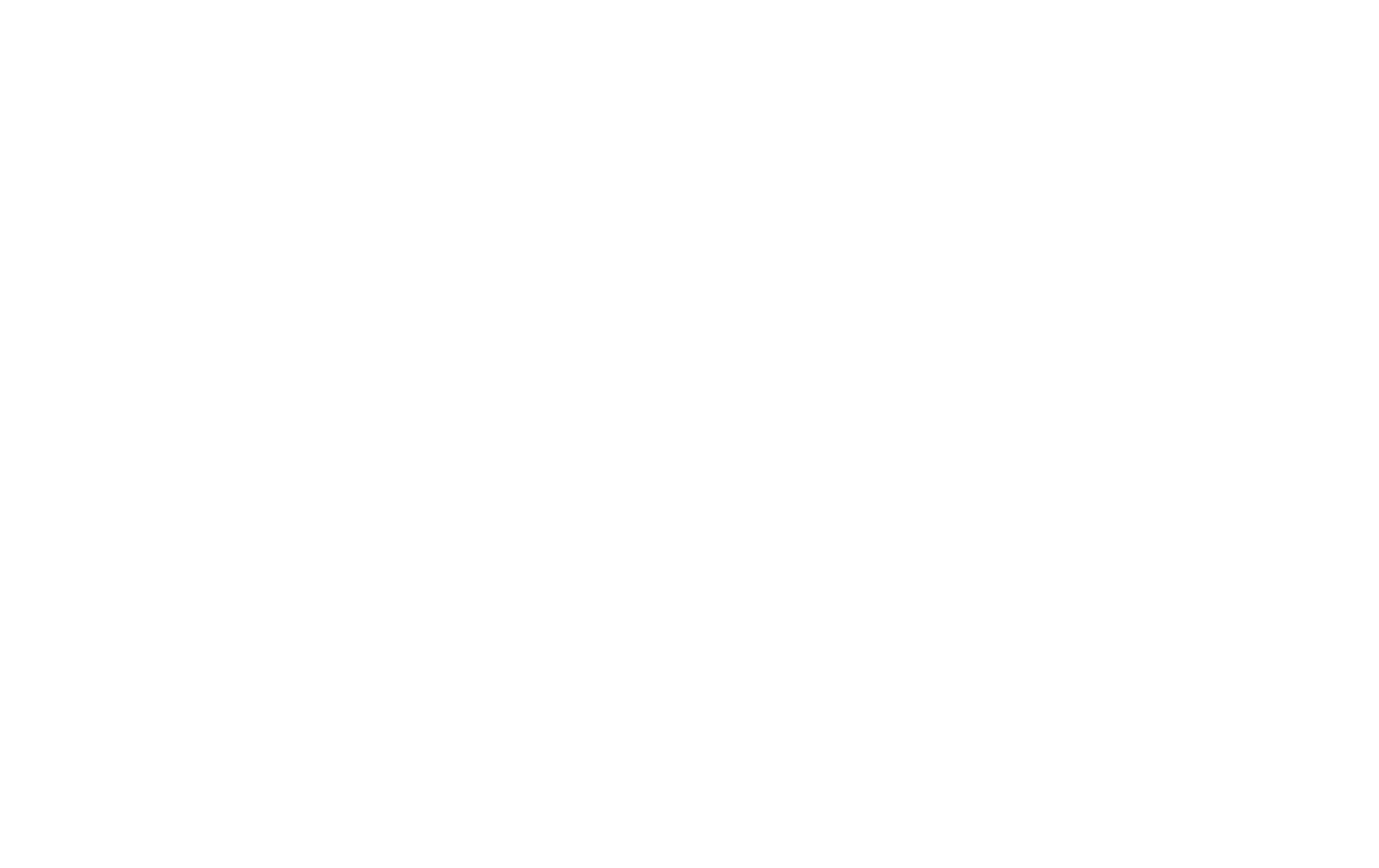}
\caption{Energy diagram for Fe II.}
\label{fig:ed}
\end{figure}

Electron-impact is a dominant populating mechanism for the excited fine-structure levels of Fe$^+$. The energy gap between the ground level of Fe$^+$ and its next four excited levels in the ground term is less than 1500 K (see Figure \ref{fig:ed}), so these fine-structure levels can be easily excited and the subsequent emission lines appear in the mid-infrared (mid-IR), therefore falling within the detector windows of many telescopes, namely the {\it Spitzer Space Telescope}, the Stratospheric Observatory for Infrared Astronomy (SOFIA), and the up-coming {\it James Webb Space Telescope (JWST)}. \cite{neufeld2007} reported the detection of \ion{Fe}{ii} (25.99 $\mu$m and 35.35 $\mu$m) emissions in supernova remnants. \cite{perlman2007} detected the  \ion{Fe}{ii} (25.99 $\mu$m) emission from M87 (the dominant galaxy in the Virgo Cluster). \cite{green2010} reported the detection of \ion{Fe}{ii} (25.99 $\mu$m) emission in the proto-stellar outflow GGD 37. More recently, \cite{harper2017} investigated SOFIA-EXES mid-IR observations of forbidden \ion{Fe}{ii} emissions in the early-type M super-giants and spectrally resolved the \ion{Fe}{ii} (25.99 $\mu$m) emission line from Betelgeuse. 

The \ion{Fe}{ii} fine-structure lines can serve as diagnostics of the local physical conditions of many cool plasma environments. For example, solving for the thermal balance and chemistry self-consistently, \cite{gorti2004} modeled the IR spectra from intermediate-aged disks around G and K stars and found that the \ion{Fe}{ii} (25.99 $\mu$m) emission is among the strongest features. Assuming thermal pressure balance, \cite{kaufman2006} calculated \ion{Fe}{ii} (25.99 $\mu$m) emission that may arise from \ion{H}{ii} regions and/or photodissociation regions (PDRs) in massive star-forming environments. Of objects for which local thermodynamic equilibrium (LTE) is not valid, the physical conditions can be extracted from the spectra only when the collisional rates are known. For example, \cite{bautista1996excitation} studied the excitation of \ion{Ni}{ii} and \ion{Fe}{ii} based on collisional data from \cite{bautista1996atomic}. \cite{verner1999} performed numerical simulations of \ion{Fe}{ii} emission spectra with the simulation code CLOUDY using the collision strengths from \cite{zhang1995}. The same data set was also used by \cite{hartigan2004} to model \ion{Fe}{ii} (25.99 $\mu$m) emission as a diagnostic of shocked gas in stellar jets and more recently by \cite{lind2017} to study non-LTE line formation of Fe in late-type stars. 

There has been considerable effort and resources dedicated to the computation of electron-impact fine-structure excitation rates for Fe$^+$. Such calculations are challenging and demanding for many reasons. First, the competition between filling of the $3d$ and $4l$ shells makes it a non-trivial exercise to obtain a sufficiently accurate atomic structure target model. Second, the number of closely coupled channels increases dramatically as more configurations are added to the target model. A compromise has to be made between accuracy and computational resources. Third, the existence of Rydberg resonance series below each excitation threshold requires a very fine energy mesh in order to obtain reliable effective collision strengths. 

Previous calculations can be naturally divided into two groups according to the different choice of target models. The first group considers only even-parity configurations. \citet{nussbaumer1980} calculated electron-ion collision strengths for the lowest four terms of Fe$^+$ with $3d^64s$ and $3d^7$ in the target model. \citet{berrington1988} and \citet{keenan1988} extended this work by including a $\bar{4}d$ pseudo-orbital and applying the BP approximation. Recently \citet{bautista2015} reported new effective collision strengths applying the ICFT $R$-matrix method \citep{griffin1998} and the Dirac Atomic $R$-matrix code (DARC) \citep{norrington1981,dyall1989,norrington2004}. For the excitation from the ground level to the first excited level, most of their ICFT calculations give $\Upsilon$ (10$^4$~ K) of about 2, while their DARC calculation gives $\Upsilon$ (10$^4$~K) about 5. However, the electron configurations in their ICFT target are all of even parity, while the target for DARC contained $3d^64p$. The usage of different target models as well as $R$-matrix method makes it difficult to attribute the variation of the collision results to a particular reason. 

The second group of calculations considered the $3d^64p$ configuration in the target model. \cite{pradhan1993b} carried out two sets of close-coupling calculations. The first set included 38 quartet and sextet terms belonging to the $3d^64s$, $3d^7$ and $3d^64p$ using the non-relativistic (NR) $LS$ coupling $R$-matrix package  \citep{berrington1987} with $\bar{4}d$ correlation orbital included in some of the configurations. The second set was carried out using the semi-relativistic Breit-Pauli $R$-matrix package \citep{scott1982} with only 41 fine-structure levels included, primarily due to the limit of computation capability. These $LS$ coupling calculations were then extended by \citet{pradhan1993z} and \citet{zhang1995} to obtain fine-structure effective collision strengths using a recoupling method. \citet{ramsbottom2007} presented new fine-structure calculations using a parallel Breit-Pauli (BP) $R$-matrix package \citep{ballance2004}. Their target model contained $3d^64s$, $3d^7$, $3d^64p$ with additional correlation effects incorporated via the $3d^6\bar{4}d$ configuration. In addition, \citet{bautista1996atomic} and \citet{bautista1998} studied the influence of including doublets arising from $3d^54s^2$ and found the collision strengths of the $^6D_{9/2}$-$^6D_{7/2}$ transition have similar background values to those without doublets. On average, calculations from the second group tend to give similar effective collision strengths to each other, and show larger differences with those from the first group. 

The primary aim of this paper is to evaluate accurate low temperate rate coefficients for fine-structure transitions within the ground term of \ion{Fe}{ii}, while previous work primarily focused on high temperatures ($\geq 2,000$ K).  Low-temperature collision data, required for cool plasma environments, can only be extrapolated from the available high-temperature data, which would inevitably generate large uncertainties and compromise the reliability of astronomical spectra analysis. A second aim of this work is elucidate the reason for the inconsistency reported by \citet{bautista2015}. Similar work was performed by \cite{badnell2014} who discussed the differences (about a factor of three) between ICFT and DARC calculations for Fe$^{2+}$ reported by \cite{bautista2010}. Excellent agreement (<5\%) was found, when the exact same atomic structure and the same close-coupling expansion were adopted. Third, \citet{ramsbottom2015} presented an in-depth comparison of collision strengths and effective collision strengths produced using all variants of the $R$-matrix codes. For the selection of ions, namely \ion{Cr}{ii}, \ion{Mn}{v} and \ion{Mg}{viii}, which are important iron-peak species, the relativistic and LS transformed $R$-matrix approaches all produce rates of a similar accuracy. We will check if this conclusion is also valid for \ion{Fe}{ii}. Finally, we want to investigate the sensitivity of the effective collision strengths to the choice of target models as well as the adopted $R$-matrix method. This will allow us to evaluate the uncertainties of our results.

The rest of this paper is structured as follows. In Section \ref{sec:theory} we provide a brief guide to the three (ICFT, BP, and DARC) $R$-matrix methods used in this work. In Section \ref{sec:target} we built several target models and discuss the results of the atomic structure calculations. In Section \ref{sec:scattering}, we gave details of six independent $R$-matrix calculations. Collision strengths and effective collision strengths are presented in Section \ref{sec:discussion}. The reliability of the methods and rationale for choosing recommended effective collision strengths, including uncertainty estimates, are also addressed. Our findings are summarized in Section \ref{sec:summary}.

%%%%%%%%%%%%%%%%%%%%%%%%%%%%%%%%%%%%%%%%%%%%%%%%%%%%%%%%%%%%%%%%%%%%%%%%%%
%%%%%%%%%%%%%%%%%%%%collision theory%%%%%%%%%%%%%%%%%%%%%%%%%%%%%%%%%
%%%%%%%%%%%%%%%%%%%%%%%%%%%%%%%%%%%%%%%%%%%%%%%%%%%%%%%%%%%%%%%%%%%%%%%%%%%
\section{collision Theory}
\label{sec:theory}

The theory behind the $R$-matrix method has been well documented in the literature \citep{eissner1974,hummer1993,burke2011} and many versions of computer packages adopting the $R$-matrix approach have been developed in the past decades. Generally speaking, for collisions involving heavy atoms, the relativistic effects are expected to be important and have to be included in the scattering calculation. In the DARC $R$-matrix collision program, the relativistic effects are introduced via the Dirac Hamiltonian. In another commonly used semi-relativistic BP $R$-matrix (BPRM) approach, one-body relativistic terms (relativistic mass-correction, one-electron Darwin, and the spin-orbit term) are considered in the Hamiltonian. Good agreement between these two methods was found by \cite{berrington2005} in the study of  Fe$^{14+}$ in collision with electrons, when target states of the two methods are in agreement and resonances are resolved adequately.

However, computational challenges arises when using the DARC or BPRM approaches. The inclusion of the spin-orbit term requires $jj$ (for DARC) and $jK$ (for BP) coupling; the size of the Hamiltonian matrices that need to be diagonalized can become very large. Many frame-transformation methods have been developed to make the calculations possible and less time-consuming. One such procedure is called intermediate coupling frame transformation (ICFT) \citep{griffin1998}. The ICFT $R$-matrix approach carries only the non-fine-structure terms, mass-correction and Darwin terms in the Hamiltonian operators of the interior region. The spin-orbit term is only considered in the Hamiltonian operators for the exterior and asymptotic regions. On the boundary, multi-channel quantum defect theory (MQDT) is employed to generate $LS$-coupled `unphysical' $K$-matrices  and those matrices are then transformed into a $jK$ coupling representation. Since the Hamiltonian matrix in the interior region is written in $LS$ coupling, the diagonalization of the ICFT $R$-matrix method is an order of magnitude more efficient than BPRM. 

In this work we use all three $R$-matrix methods described above, with the calculations for each one optimized for the case of fine-structure excitation within the ground term of \ion{Fe}{ii}.

%%%%%%%%%%%%%%%%%%%%%%%%%%%%%%%%%%%%%%%%%%%%%%%%%%%%%%%%%%%%%%%%%%%%%%%%%%
%%%%%%%%%%%%%%%%%%%%target model%%%%%%%%%%%%%%%%%%%%%%%%%%%%%%%%%
%%%%%%%%%%%%%%%%%%%%%%%%%%%%%%%%%%%%%%%%%%%%%%%%%%%%%%%%%%%%%%%%%%%%%%%%%%%
\begin{table}
\caption{Breit-Pauli/ICFT target models for \ion{Fe}{ii}.}
\begin{tabular}{cccc}
\hline \hline
Model                &3-even                &3-mix                 &4-mix\\
\hline
Target               &$3d^64s$, $3d^7$      &$3d^64s$, $3d^7$      &$3d^64s$, $3d^7$ \\
                     &$3d^54s^2$            &$3d^64p$              &$3d^64p$, $3d^64d$\\
\hline
Scaling              &$\lambda_{1s}$=1.0000 &$\lambda_{1s}$=1.0000 &$\lambda_{1s}$=1.00000\\
parameter           &$\lambda_{2s}$=0.9000 &$\lambda_{2s}$=0.9000 &$\lambda_{2s}$=1.27407\\
                     &$\lambda_{2p}$=1.0360 &$\lambda_{2p}$=1.0360 &$\lambda_{2p}$=1.11361\\
                     &$\lambda_{3s}$=1.1000 &$\lambda_{3s}$=1.1000 &$\lambda_{3s}$=1.09525\\
                     &$\lambda_{3p}$=1.0050 &$\lambda_{3p}$=1.0050 &$\lambda_{3p}$=1.05904\\
                     &$\lambda_{3d}$=1.0381 &$\lambda_{3d}$=1.0381 &$\lambda_{3d}$=1.04657\\
                     &$\lambda_{4s}$=0.9400 &$\lambda_{4s}$=0.9400 &$\lambda_{4s}$=0.89000\\
                     &                      &$\lambda_{4p}$=0.8000 &$\lambda_{4p}$=0.98955\\
                     &                      &                      &$\lambda_{4d}$=1.34726\\
\hline
target levels        &119                   &262                   &538\\
target terms         &48                    &100                   &204\\
\hline
$(N+1)$             &$3d^8$                &$3d^8$                &$3d^8$\\       
bound               &$3d^74s$              &$3d^7\{4s,4p\}$       &$3d^7\{4s,4p,4d\}$ \\
system              &$3d^64s^2$            &$3d^6\{4s^2,4p^2\}$   &$3d^6\{4s^2,4p^2,4d^2\}$ \\
                    &                      &$3d^64s4p$            &$3d^6\{4s4p,4s4d,4p4d\}$ \\
\hline
RA(BP)  & 12.11523      &  18.17773    &16.92773 \\
RA(ICFT)& 12.86523      &  18.17773    &-\\
\hline\hline
\end{tabular}
\label{tab:target}

\raggedright \textbf{Notes.} RA (in units of a.u.) represents the $R$-matrix boundary.
\end{table}

\begin{table*}
\caption{Level energies (in Ry) of Fe$^+$.}
\begin{tabular}{ccc cc cc cc cc}
\hline \hline
No. &Term/Level    &Observed$^a$ &GRASP$^0$            &4-mix             &3-mix             &3-even            &R07$^b$      &B88$^b$\\
\hline
 1&$3d^64s$ $^6D_{9/2}$&0.000000 &0.000000( 0.00\%) &0.000000( 0.00\%) &0.000000( 0.00\%) &0.000000( 0.00\%) &0.00000 &0.00000\\
 2&$3d^64s$ $^6D_{7/2}$&0.003506 &0.003226( 7.99\%) &0.003624( 3.37\%) &0.003479( 0.77\%) &0.003479( 0.77\%) &0.00398 &0.00324\\
 3&$3d^64s$ $^6D_{5/2}$&0.006084 &0.005620( 7.63\%) &0.006350( 4.37\%) &0.006091( 0.12\%) &0.006091( 0.12\%) &0.00695 &0.00560\\
 4&$3d^64s$ $^6D_{3/2}$&0.007861 &0.007278( 7.42\%) &0.008251( 4.96\%) &0.007913( 0.66\%) &0.007913( 0.66\%) &0.00903 &0.00725\\
 5&$3d^64s$ $^6D_{1/2}$&0.008904 &0.008253( 7.31\%) &0.009376( 5.30\%) &0.008990( 0.97\%) &0.008990( 0.97\%) &0.01025 &0.00822\\
\hline
 6&$3d^7$ $^4F_{9/2}$  &0.017064 &0.017568( 2.95\%) &0.017828( 4.48\%) &0.016460( 3.54\%) &0.016118( 5.54\%) &0.02898 &0.05030\\
 7&$3d^7$ $^4F_{7/2}$  &0.022145 &0.021970( 0.79\%) &0.022420 (1.24\%) &0.022282( 0.62\%) &0.021939( 0.93\%) &0.03428 &0.05540\\
 8&$3d^7$ $^4F_{5/2}$  &0.025862 &0.025231( 2.44\%) &0.025843( 0.07\%) &0.026624( 2.95\%) &0.026279( 1.61\%) &0.03820 &0.05930\\
 9&$3d^7$ $^4F_{3/2}$  &0.028409 &0.027481( 3.27\%) &0.028214( 0.69\%) &0.029632( 4.30\%) &0.029286( 3.09\%) &0.04091 &0.06210\\
\hline
10&$3d^64s$ $^4D_{7/2}$&0.072494 &0.106004(46.22\%) &0.052940(26.97\%) &0.071998( 0.68\%) &0.071893( 0.83\%) &0.07233 &0.09110\\
11&$3d^64s$ $^4D_{5/2}$&0.076473 &0.108595(42.00\%) &0.057164(25.25\%) &0.076053( 0.55\%) &0.075949( 0.69\%) &0.07696 &0.09470\\
12&$3d^64s$ $^4D_{3/2}$&0.079102 &0.109831(38.85\% )&0.060015(24.13\%) &0.078798( 0.38\%) &0.078694( 0.52\%) &0.08007 &0.09730\\
13&$3d^64s$ $^4D_{1/2}$&0.080618 &0.110623(37.22\%) &0.061673(23.50\%) &0.080396( 0.28\%) &0.080293( 0.40\%) &0.08187 &0.09890\\
\hline
14&$3d^7$ $^4P_{5/2}$  &0.122788 &0.112219( 8.61\%) &0.139806(13.86\%) &0.175247(42.72\%) &0.174719(42.29\%) &0.15040 &0.17180\\
15&$3d^7$ $^4P_{3/2}$  &0.124599 &0.112359( 9.82\%) &0.141839(13.84\%) &0.177758(42.66\%) &0.177239(42.25\%) &0.15270 &0.17480\\
16&$3d^7$ $^4P_{1/2}$  &0.126710 &0.113633(10.32\%) &0.143726(13.43\%) &0.180416(42.38\%) &0.179894(41.97\%) &0.15492 &0.17660\\
\hline
  & Averaged Error&         &13.17\%           &9.73\%            &8.45\%            &8.39\%            &        &       \\
  \hline
  ...\\
\hline 
64&$3d^64p$ $^6D^o_{9/2}$&0.350464 &0.353878( 0.97\%) &0.262281(25.16\%) &0.467892(33.51\%) & -       &0.32084 & -\\
65&$3d^64p$ $^6D^o_{7/2}$&0.352296 &0.355631( 0.95\%) &0.264744(24.85\%) &0.470532(33.56\%) & -       &0.32334 & -\\
66&$3d^64p$ $^6D^o_{5/2}$&0.354109 &0.357276( 0.89\%) &0.266864(24.64\%) &0.472772(33.51\%) & -       &0.32557 & -\\
67&$3d^64p$ $^6D^o_{3/2}$&0.355515 &0.358547( 0.85\%) &0.268450(24.49\%) &0.474447(33.45\%) & -       &0.32725 & -\\
68 &$3d^64p$ $^6D^o_{1/2}$&0.356390 &0.359340( 0.68\%) &0.269425(24.52\%) &0.475475(33.21\%) & -       &0.32830 & -\\
\hline\hline
\end{tabular}
\label{tab:level}\\
\raggedright Notes: $^a$ \citet{NIST}.  $^b$ R07 = \citet{ramsbottom2007} and B88= \citet{berrington1988}.\\
\end{table*}

\begin{table*}
\caption{Einstein A coefficient (in $s^{-1}$) for \ion{Fe}{ii}.}
\begin{tabular}{ccccccc}
\hline \hline
Transition&$^6D_{9/2}-^6D_{7/2}$&$^6D_{7/2}-^6D_{5/2}$&$^6D_{5/2}-^6D_{3/2}$&$^6D_{3/2}-^6D_{1/2}$&$^6D_{9/2}-^6D_{9/2}^o$\\\\
Type & M1 &M1&M1&M1&E1\\
Wavelength($\mu$m)&25.988&35.349&51.301&87.384&259.940\\
\hline
NIST$^a$    &2.13               &1.57               &7.19               &1.89              &2.35\\ 
3-even      &2.084( 2.16\%)     &1.632( 3.95\%)     &7.750( 7.79\%)     &2.076( 9.84\%)    &-\\
3-mix       &2.084( 2.16\%)     &1.632( 3.95\%)     &7.750( 7.79\%)     &2.076( 9.84\%)    &7.634 \\
4-mix       &2.355(10.56\%)     &1.853(16.88\%)     &8.816(22.60\%)     &2.364(25.08\%)    &1.649  \\
DARC        &1.662(21.97\%)     &1.256(20.00\%)     &5.834(18.86\%)     &1.542(18.41\%)    &2.53 \\
Q$^b$           &$10^{-3}$ &$10^{-3}$ &$10^{-4}$ &$10^{-4}$ \\
\hline \hline
\end{tabular}
\label{tab:Avalue}

\raggedright \textbf{Notes.}  $^a$ \citet{NIST}. $^b$ Units of  $Q\times s^{-1}$.
\end{table*}

\section{target model}
\label{sec:target}
For fine-structure excitation of \ion{Fe}{ii} within the ground term, low-energy electron-ion collisions are dominated by resonance structures. 
If the ionic states themselves are not accurately represented in the target model, this inaccuracy will affect the collision strengths by shifting the resonance peaks to wrong positions. However, obtaining a sufficiently accurate atomic structure for \ion{Fe}{ii} is a non-trivial exercise. 

The atomic structure program AUTOSTRUCTURE \citep{badnell1997,badnell2011} includes one-body relativistic corrections and was used to generate targets for the BP and ICFT $R$-matrix methods. We built three small-scale target models (see Table \ref{tab:target}). What we refer to as the 3-even target model contains only even-parity configurations, so that the following scattering calculations exclude dipole transitions, which should be much stronger than the fine-structure transitions in which we are interested. However, as mentioned in Section \ref{sec:introduction}, it was found that the $3d^64p$ configuration played a vital role in the transitions among the low-lying fine-structure levels, which is possibly due to its coupling with $3d^64s$. In the 3-mix target model, we include the $3d^64p$ configuration and the same scaling parameters as for target 3-even. The spectroscopic configuration $3d^64d$ is retained in the 4-mix target model.

While there is an iterative variational procedure implemented in AUTOSTRUCTURE, satisfactory level energies of the first excited term $a ^4F$ cannot be obtained without the inclusion of $4d$ orbitals. To improve the target structure further, we developed a code to vary scaling parameters associated with the Thomas-Fermi-Dirac-Amaldi potential and then compared the resulting energies until a minimum was found in the differences with the NIST \citep{NIST} level energies. In the code that was developed for this optimization, a grid of $\lambda_{nl}$ parameters was chosen, followed by a comparison with NIST level energies for the levels of the ground term. A subset of these, which gave the closest agreement with NIST, was then chosen and the level energies of the first excited term. A subset of these was then examined, comparing the level energies of the higher excited levels. In addition, a comparison with NIST A-values for the transitions within ground term was also performed, to sub-select on the the set of $\lambda_{nl}$ that were closest to NIST A-values. It was found that this method gave better agreement for the energies of the low lying terms and associated A-values that the existing optimization procedure within AUTOSTRUCTURE, when it was optimized on just the first few terms. This variation method is used for 3-even and 3-mix target models, and the built-in AUTOSTRUCTURE variation procedure is used for 4-mix target model.

The target model for the DARC calculation was obtained via the multi-configuration Dirac-Fock method using the computer package GRASP$^0$ \citep{dyall1989,parpia1996}. The Fe$^+$ target has been investigated by \cite{smyth2018}. We adopt their 20 configuration target model: $3d^7$; $3d^6{4s,4p,4d,5s,5p}$; $3d^5{4s^2,4p^2,4d^2,4s4p,4s4d,5s^2,5p^2}$; $3p^4{3d^9,4d^9}$; $3p^6ed4d^6$; $3p^63d^24d^5$; and $3p^63d^34d^4$. The full DARC target gives 6069 levels. Extended Average Level (EAL) optimization option were used in the GRASP$^0$ structure calculation to optimize the level energies upon all of the levels. Better overall atomic structure should be obtained through this method.

A selection of fine-structure level energies are presented in Table \ref{tab:level}. Compared with previous work \citep{berrington1988,ramsbottom2007}, significant improvements are achieved in the first 9 levels (term $^6D$ and $^4F$). Target 4-mix estimates $^4P$ term/level energies better than the target 3-mix and 3-even models, but gives worse $^4D$ term/level energies. None of the three small-scale BP/ICFT targets can predict the $^6D^o$ term/level energies well, mainly because of the limited target size. The DARC target gives very good $^6D^o$ term/level energies, but overestimates the $^4D$ term/level energies. 

Radiative rates (A-values) for fine-structure transitions in the ground term as well as the first dipole transition are presented  in Table \ref{tab:Avalue} and compared to NIST values \citep{NIST}. The three AUTOSTRUCTURE targets give better M1 transitions, while the DARC target gives better E1 transitions. The difference in the results comes from the different target wavefunctions and computational methods. In our GRASP$^0$ calculation, energies are optimized upon all of the level energies included in the calculation, so better overall atomic structure should be obtained, while the lowest few levels are not so well optimized as compared to our AUTOSTRUCTURE targets. We believe that this explains some of the differences between the GRASP$^0$ and NIST M1 A-values. As will be shown later, except for the 3-even calculation all other targets give similar collision strengths. So the final results are not highly sensitive to the A-value differences. 

%%%%%%%%%%%%%%%%%%%%%%%%%%%%%%%%%%%%%%%%%%%%%%%%%%%%%%%%%%%%%%%%%%%%%%%%%%
%%%%%%%%%%%%%%%%%%%%scattering calculation%%%%%%%%%%%%%%%%%%%%%%%%%%%%%%%%%
%%%%%%%%%%%%%%%%%%%%%%%%%%%%%%%%%%%%%%%%%%%%%%%%%%%%%%%%%%%%%%%%%%%%%%%%%%%
\section{scattering calculation}
\label{sec:scattering}
We performed two ICFT $R$-matrix calculations with the 3-even and 3-mix targets, three BP $R$-matrix calculations with 3-even, 3-mix and 4-mix targets, and one DARC calculation. In the 3-mix BP, 4-mix BP, and DARC calculations, the full configuration target was taken through until the Hamiltonian diagonalization, and then the first 100 levels are shifted to NIST values and retained in the rest of the calculation. This process will still include all of the important resonance contributions to the fine-structure excitations reported on in the paper, given the low temperature focus of this work. 

The scattering calculation included $J\Pi$ partial waves from $2J=0$ to $2J$ = 30 with 20 continuum basis terms for each value of angular momentum. Total angular momenta $L\leq 18$ and $1\leq(2S+1)\leq 7$ were used for both the even and odd parities. The contributions from higher $J$ were obtained from the top-up procedure. The $R$-matrix boundaries for the different collision calculations were automatically selected by the $R$-matrix code. The ($N$+1) bound configurations included in the scattering computations are listed targets in Table \ref{tab:target}. Convergence checks on the size of the continuum basis used was determined by identifying the most dominant partial waves ($2J=$ 8 and 10). We found that 15 continuum basis functions were sufficient to ensure convergence for the fine-structure transitions of interest. 

Collision strengths are sampled using a very fine energy mesh of $2.5\times 10^{-5}$ Ryd up to 0.1035 Ry and then $10^{-4}$ Ryd up to 0.6035 Ry. Coarse meshes with an interval of $10^{-3}$ Ryd with different numbers of energy points are tested up to 2.6035 Ryd. Adding more data points within the coarse mesh doesn't show any noticeable differences in final effective collision strengths.

%%%%%%%%%%%%%%%%%%%%%%%%%%%%%%%%%%%%%%%%%%%%%%%%%%%%%%%%%%%%%%%%%%%%%%%%%%
%%%%%%%%%%%%%%%%%%%%results and discussion%%%%%%%%%%%%%%%%%%%%%%%%%%%%%%%%%
%%%%%%%%%%%%%%%%%%%%%%%%%%%%%%%%%%%%%%%%%%%%%%%%%%%%%%%%%%%%%%%%%%%%%%%%%%%
\section{results and Discussion}
\label{sec:discussion}
%%%%figures and table%%%%%%%%%%%%%%%%%%%%%%%%%%%%%%%%
\begin{figure}
\centering
\includegraphics[width=\linewidth]{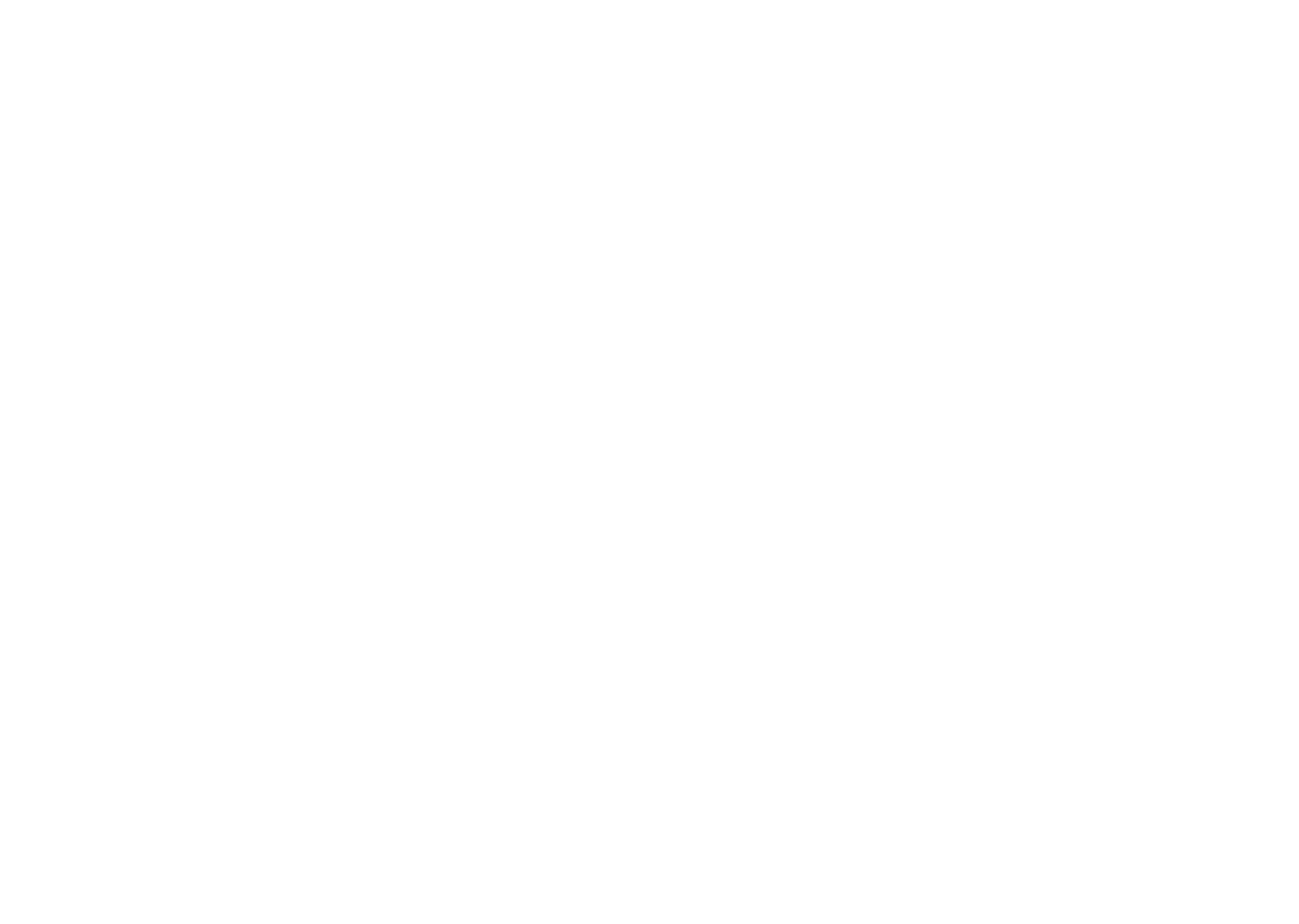}
\caption{Collision strengths for the $3d^64s$  $^6D_{9/2}$ - $3d^64s$ $^6D_{7/2}$ transition. Top panel: 3-even target model + BP $R$-matrix method; middle panel: 3-even target model + ICFT $R$-matrix method; bottom panel: 3-mix target model + ICFT $R$-matrix method.}
\label{fig:set1}
\end{figure}

\begin{figure}
\centering
\includegraphics[width=\linewidth]{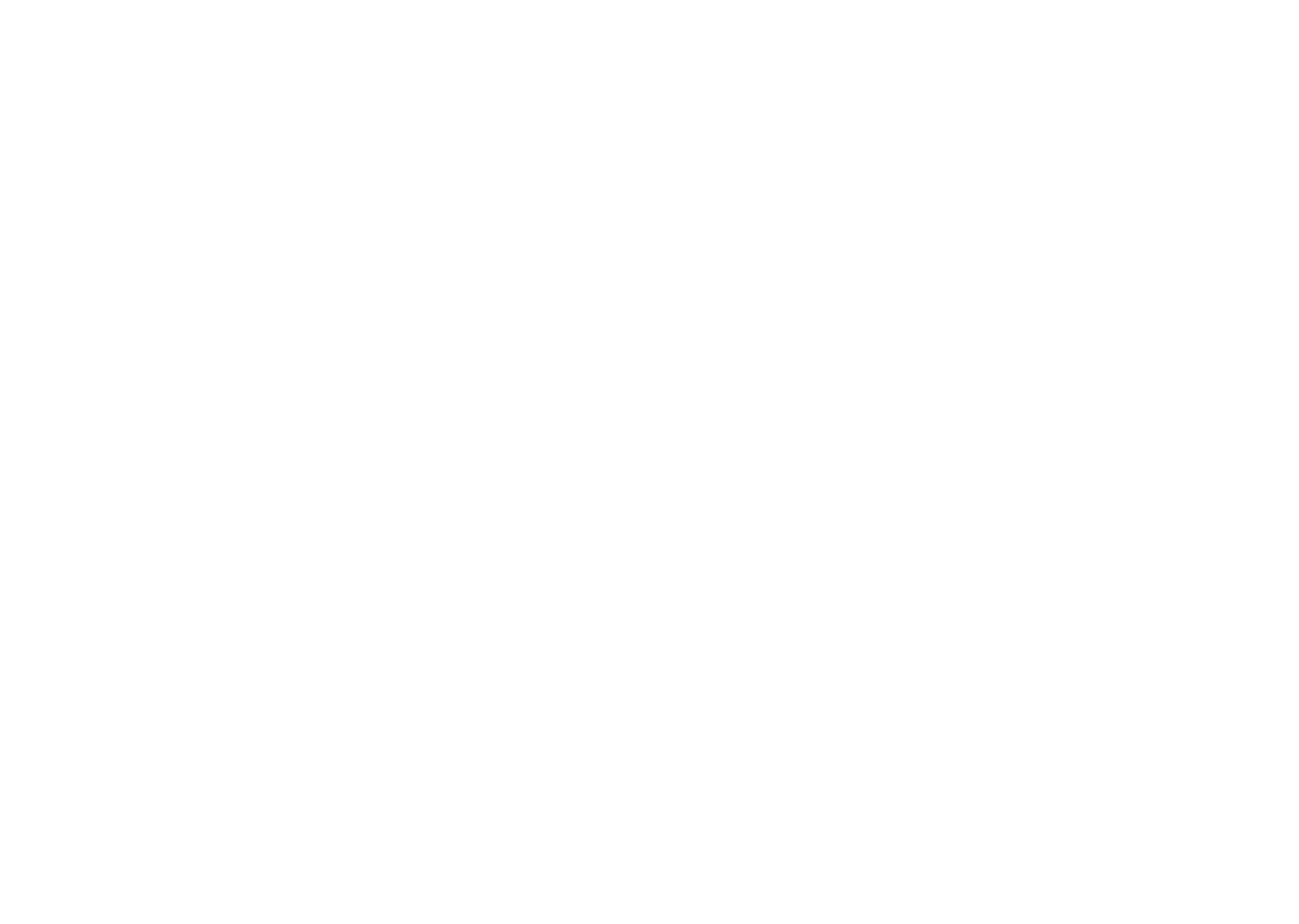}
\caption{Collision strengths for the $3d^64s$  $^6D_{9/2}$ - $3d^64s$ $^6D_{7/2}$ transition. Top panel: DARC calculation; middle panel: 4-mix target model + BP $R$-matrix method; bottom panel: 3-mix target model + BP $R$-matrix method.}
\label{fig:set2}
\end{figure}

\begin{figure}
\centering
\includegraphics[width=\linewidth]{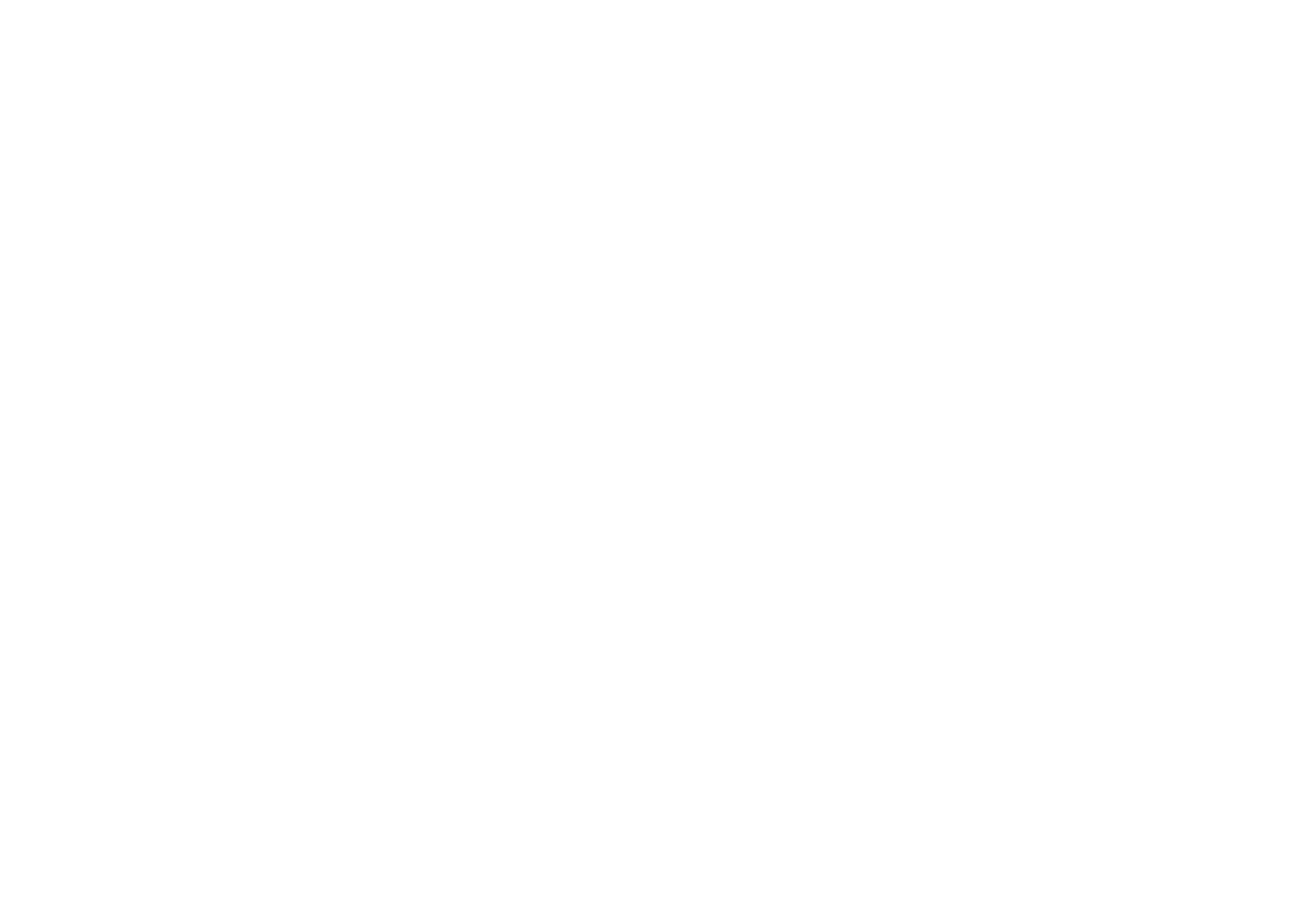}
\caption{Collision strengths for the $3d^64s$  $^6D_{9/2}$ - $3d^64s$ $^6D_{7/2}$ transition.}
\label{fig:cs1to2}
\end{figure}

\begin{figure}
\centering
\includegraphics[width=\linewidth]{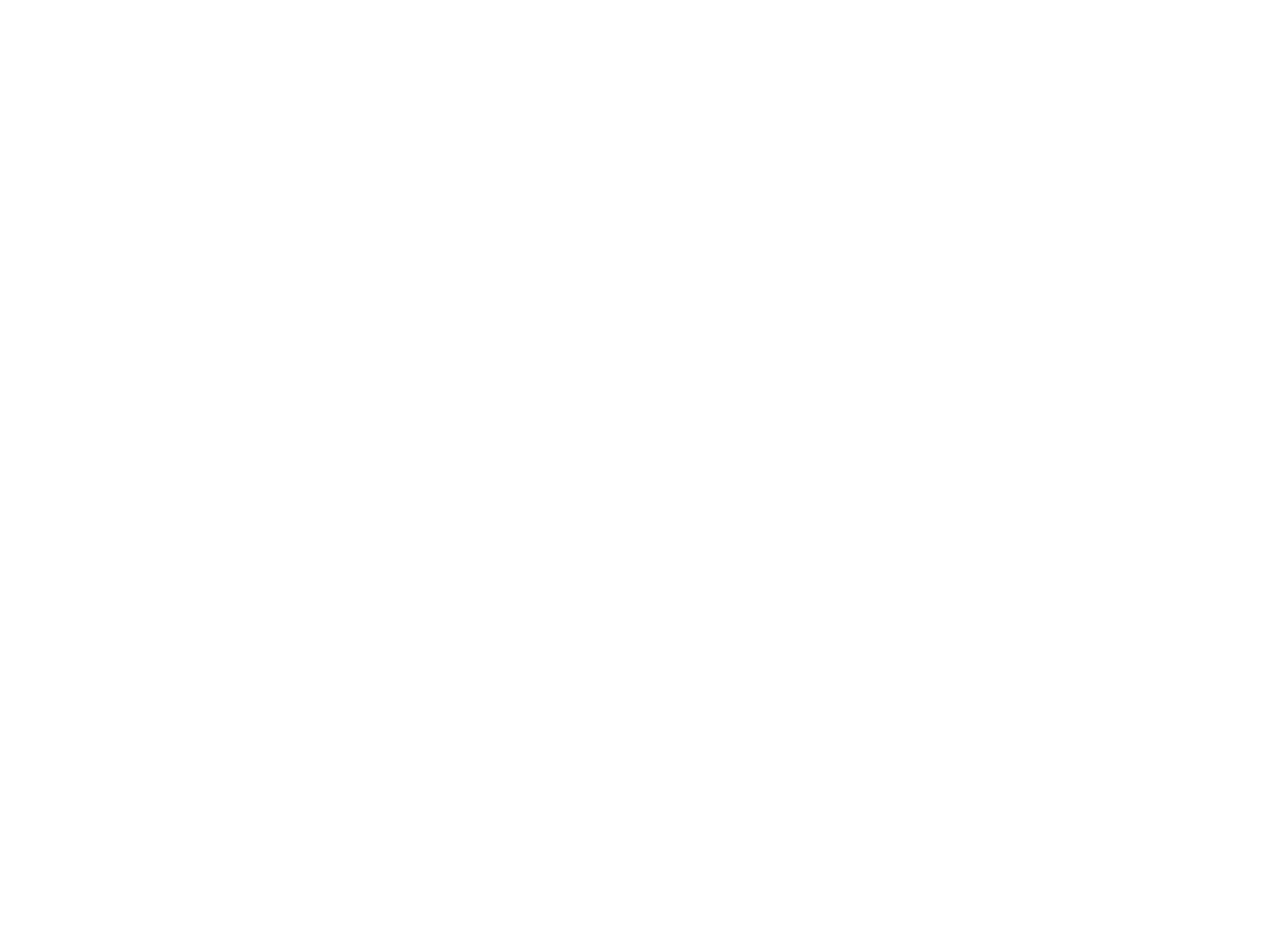}
\caption{Effective collision strengths for the $3d^64s$  $^6D_{9/2}$ - $3d^64s$ $^6D_{7/2}$ transition. Standard deviations are marked as error bars for the recommended DARC computations. Notation R07, Z95, B88 and K88 denote the results from \citet{ramsbottom2007}, \citet{zhang1995}, \citet{berrington1988} and \citet{keenan1988}, respectively.}
\label{fig:1to2}
\end{figure}

\begin{figure}
\centering
\includegraphics[width=\linewidth]{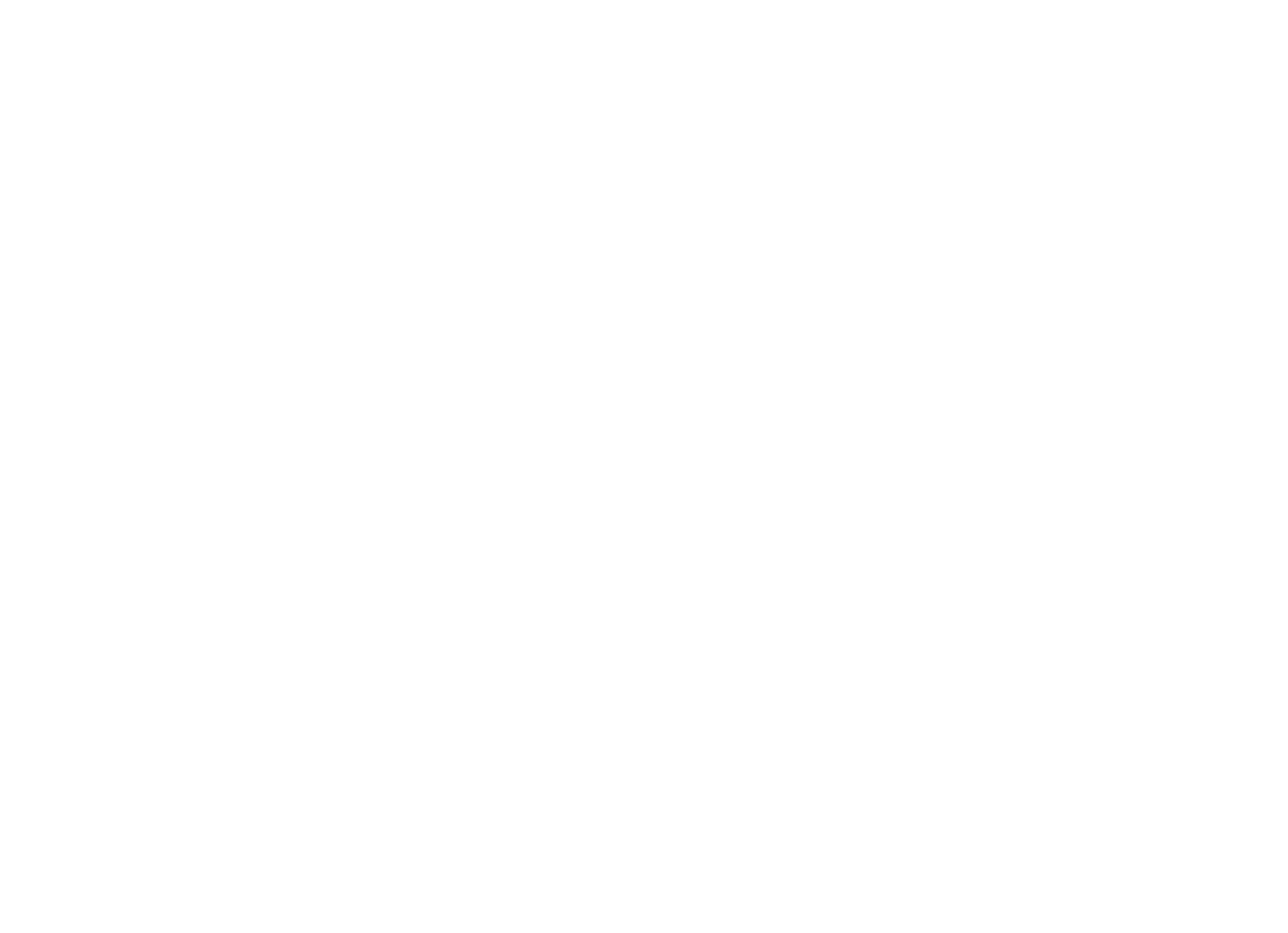}
\caption{Effective collision strengths for the $3d^64s$ $^6D_{9/2}$ - $3d^64s$ $^6D_{5/2}$ (top panel), $3d^64s$ $^6D_{9/2}$ - $3d^64s$ $^6D_{3/2}$ (middle panel) and $3d^64s$ $^6D_{9/2}$ - $3d^64s$ $^6D_{1/2}$ (bottom panel) transitions. Black circles are recommended DARC results. Standard deviations are marked as error bars. Red and blue curves are results of \citet{ramsbottom2007} and \citet{zhang1995}, respectively.}
\label{fig:from1}
\end{figure}

\begin{figure}
\centering
\includegraphics[width=\linewidth]{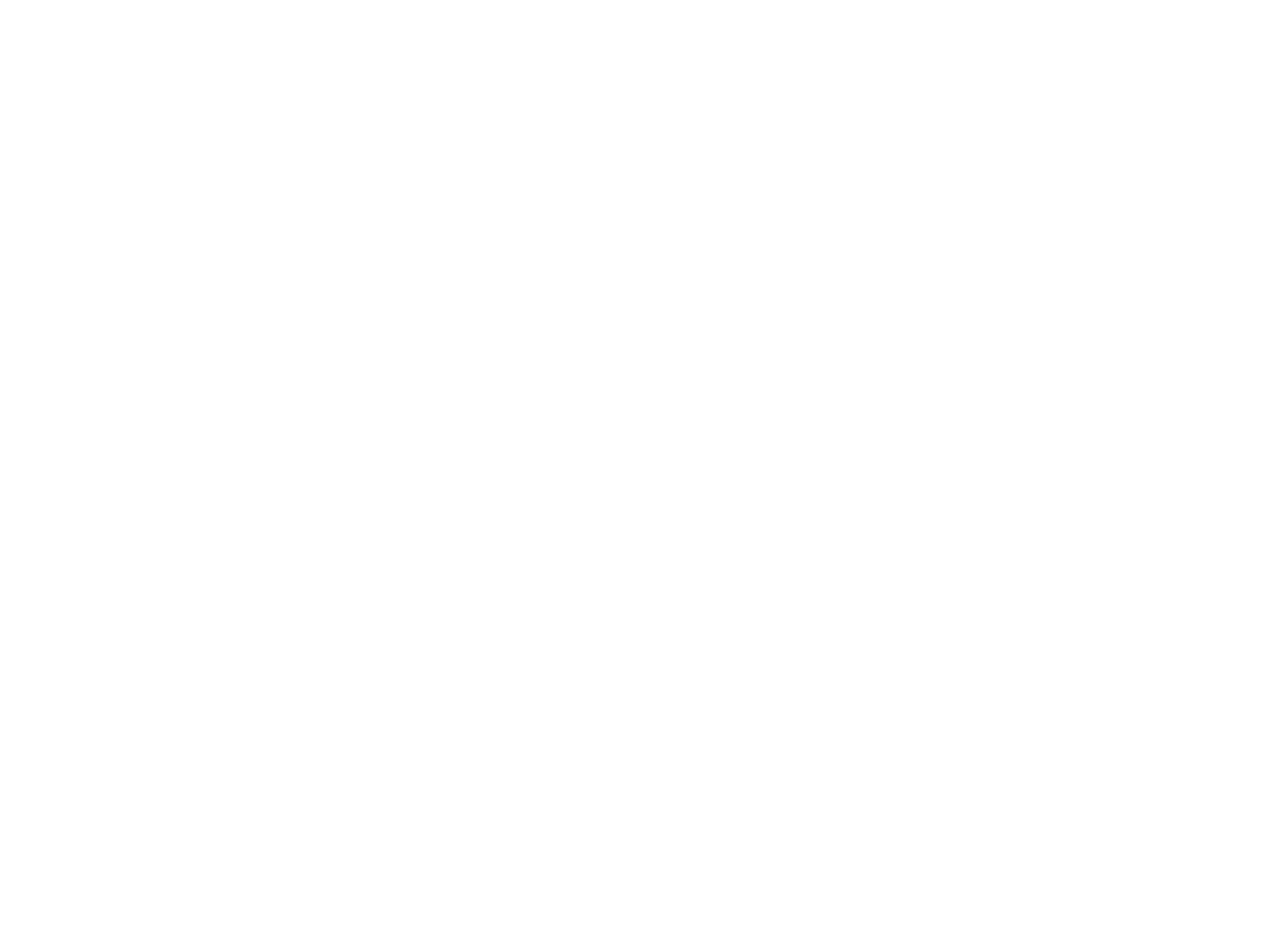}
\caption{Effective collision strengths for the $3d^64s$ $^6D_{7/2}$ - $3d^64s$ $^6D_{5/2}$ (top panel), $3d^64s$ $^6D_{9/2}$ - $3d^64s$ $^6D_{3/2}$ (middle panel) and $3d^64s$ $^6D_{9/2}$ - $3d^64s$ $^6D_{1/2}$ (bottom panel) transitions. Symbols are the same as for Figure~\ref{fig:from1}.}
\label{fig:from2}
\end{figure}

\begin{figure}
\centering
\includegraphics[width=\linewidth]{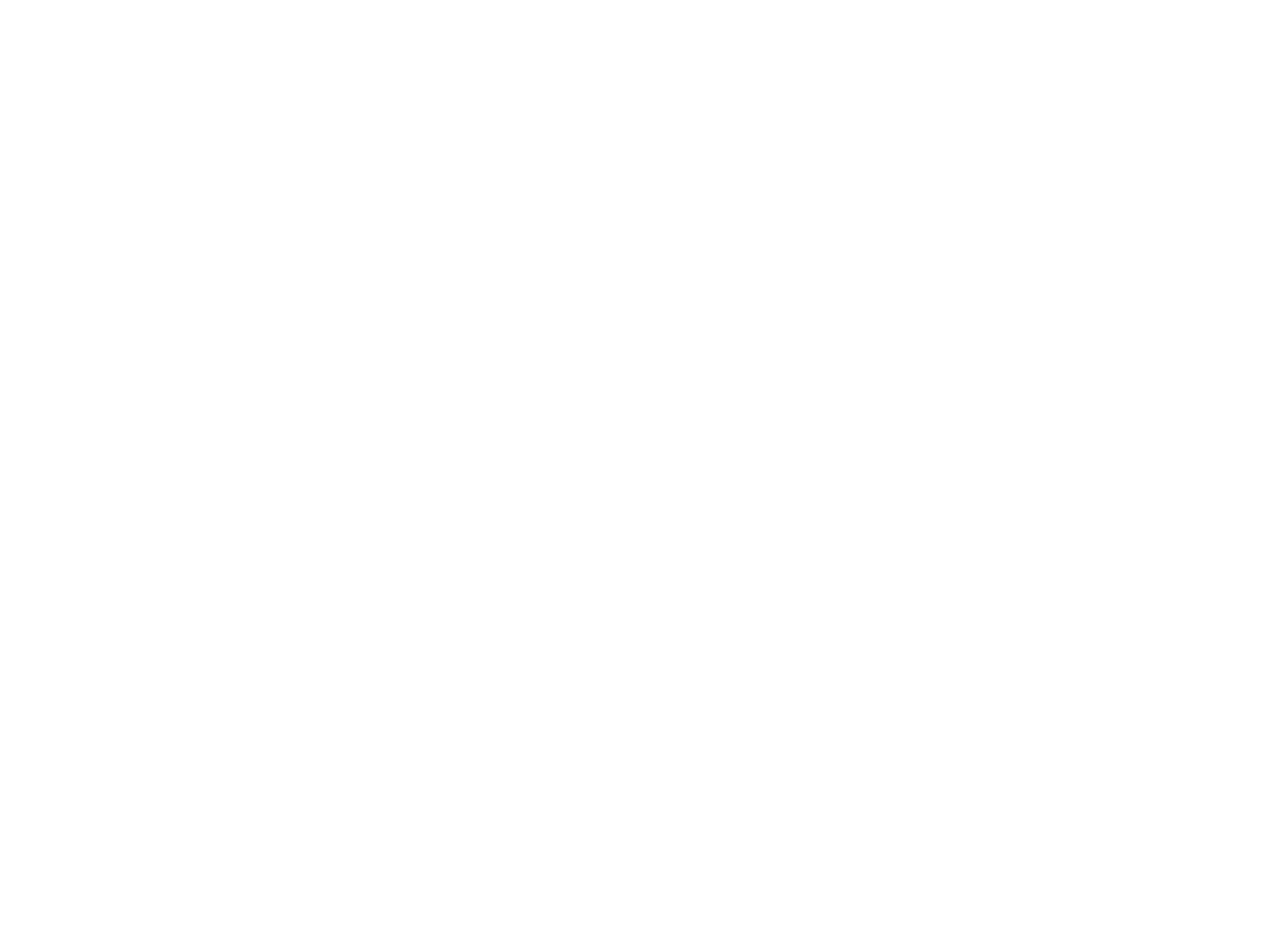}
\caption{Effective collision strengths for the $3d^64s$ $^6D_{5/2}$ - $3d^64s$ $^6D_{3/2}$ (top panel), $3d^64s$ $^6D_{5/2}$ - $3d^64s$ $^6D_{1/2}$ (middle panel) and $3d^64s$ $^6D_{3/2}$ - $3d^64s$ $^6D_{1/2}$ (bottom panel) transitions. Symbols are the same as for Figure~\ref{fig:from1}.}
\label{fig:from345}
\end{figure}

\begin{table}
\caption{Effective collision strengths for select transitions of \ion{Fe}{ii} calculated by the DARC approach. The uncertainty \% $\Delta$ was estimated by using the two BP calculations.}
\begin{tabular}{cccc}
\hline\hline
Temperature(K)&$^6D_{9/2}-^6D_{7/2}$ &$^6D_{7/2}-^6D_{5/2}$  &$^6D_{5/2}-^6D_{3/2}$ \\
\hline
 10         &3.29 (17\%) &4.92 (31\%) &3.44 (10\%)\\
 20         &3.81 (18\%) &5.60 (33\%) &3.70 (12\%)\\
100        &4.30 (20\%) &5.79 (31\%) &3.82 (13\%)\\
200        &4.09 (16\%) &5.62 (27\%) &4.08 (19\%)\\
500        &3.72 ( 7\%) &5.29 (19\%) &4.38 (24\%)\\
$10^3$     &3.85 (19\%) &5.01 ( 3\%) &4.47 (15\%)\\
$10^4$     &4.94 ( 7\%) &5.19 ( 8\%) &4.21 ( 7\%)\\
$10^5$     &4.13 ( 7\%) &3.66 ( 5\%) &2.62 ( 3\%)\\
\hline\hline
\end{tabular}
\label{tab:results}
\end{table} 
%%%%figures and table end%%%%%%%%%%%%%%%%%%%%
%%%%%%%%%%%%%%%%%%%%%%%%%%%%%%%%%%%%%%%%%%%%%

In Figures \ref{fig:set1} and \ref{fig:set2} we present the collision strengths as functions of the incident electron energy for the fine-structure transition from the ground level $3d^64s$ $^6D_{9/2}$ to the first excited level $3d^64s$ $^6D_{7/2}$. The first point to notice is that when the 3-even target is used, either the BP or ICFT collision approaches give a set of sharp resonances at energies from 0.03 to threshold (0.0089 Ryd).  The average value of the collision strengths at low energies (below 0.003 Ryd) is about 4.  This is significantly smaller than previous calculations. Second, with the inclusion of the $3d^64p$ configuration, the ICFT, BP, and DARC collision approaches all give much broader background features. Third, in the comparison of the 3-mix ICFT and 3-mix BP results, the  profiles of the two curves are generally similar except that the 3-mix ICFT results at very low energies are enhanced. In Figure \ref{fig:cs1to2}, we present the collision strengths for this transition from the current DARC, 4-mix BP and 3-mix BP calculations, but to electron impact energies as large as 0.5 Ryd. We see that there is generally good agreement between the three calculations.

The collision strength ($\Omega$) tends to vary widely from the non-resonant background value. Therefore, 
the Maxwellian averaged effective collision strength ($\Upsilon$) is preferred in astrophysics, instead of employing the collision strength. We computed the thermally averaged effective collision strengths using,
\begin{equation}
\Upsilon_{ij}(T_e)=\int_0^\infty \Omega_{ij}(E_j)\mathrm{exp}(-E_j/kT_e)\mathrm{d}(E_j/kT_e),
\label{eq:effective_cs}
\end{equation}
where $\Omega_{ij}$ is the collision strength for the transition from level $i$ to $j$. $E_j$ is the final energy of the electron, $T_e$ is the electron temperature in Kelvin and $k$ is Boltzmann's constant. It was shown to be a good approximation for a positive ion that if $\Omega$ varies with energy much more slowly than does the exponential in equation \ref{eq:effective_cs}, one may equate $\Upsilon$ to the threshold value of $\Omega$ \citep{seaton1953}. 

The effective collision strength for the transition $3d^64s$ $^6D_{9/2}$ - $3d^64s$ $^6D_{7/2}$ is presented in Figure \ref{fig:1to2}. We clearly see that applying the 3-even target, the ICFT and BP $R$-matrix approaches yield good agreement and the results are in reasonable agreement with the previous calculated values of \citet{keenan1988} and \citet{berrington1988}. Their target models only included configurations of even parity as well. The inclusion of the $3d^64p$ configuration enhances the calculated effective collision strength and thus our 3-mix, 4-mix results are more consistent with previous calculations of \citet{zhang1995} and \citet{ramsbottom2007} and with our DARC results (see below). 

As stated in Section \ref{sec:introduction}, there have been several \ion{Fe}{ii} fine-structure data sets calculated, but discrepancies exists mainly due to the different target models and $R$-matrix methods adopted.  Similar trends occur in this work as well. First, among all the target models, the GRASP$^0$ target is considered to be the best, as generally it predicts the energies of the lowest 16 levels well and for other highly excited levels it gives correct relative positions (see Table~\ref{tab:level}). Among the three models used in the BP calculations, target 3-even is not sufficient. The comparison between the 3-mix BPRM and 3-mix ICFT models shows that the ICFT approach cannot give reliable fine-structure transitions for \ion{Fe}{ii}. This may be due to the fact that the ICFT method solves the inner region problem in $LS$ coupling for a configuration-mixed target, and thus our ICFT calculations were not shifted to NIST level energies. Previous works comparing the BPRM and ICFT methods showed very good agreement between the collision cross sections and rates for both methods, when the same target description was used \citep{badnell2014}. We expect that the difference between our ICFT and BPRM results was primarily due to these differences in target energies. Thus, while we do not use the ICFT results when calculating our uncertainties, they are an indication of the likely differences between previous unshifted ICFT calculations and calculations that shifted to NIST energies. We note that shifting to target energies in an ICFT calculation is described in detail in \citet{del2014atomic}. It is evident in Figure \ref{fig:1to2} that the 3-mix BP, 4-mix BP, and DARC calculations agree overall. Therefore, we adopted the effective collision strengths from the DARC calculations as our recommended values. Results from the 3-mix BP and 4-mix BP models are used to calculate the standard deviation from the recommended values at each temperature point. We present the recommended effective collision strengths and standard deviation (marked as error bars) for the other nine fine-structure transitions within the ground term in Figures \ref{fig:from1}, \ref{fig:from2}, and \ref{fig:from345}. The results from \cite{ramsbottom2007} and \cite{zhang1995} are also plotted for comparison. Part of our results is tabulated in Table \ref{tab:results}.

\section{summary}
\label{sec:summary}
In this work we studied the electron-impact fine-structure excitation of \ion{Fe}{ii}. Two ICFT calculations with the 3-even and 3-mix targets, three BPRM calculations with 3-even, 3-mix and 4-mix targets, and one DARC calculation based on a reliable 20 configuration atomic structure model were tested and small-scale computations were performed. The full configuration target was taken through until the Hamiltonian diagonalization, and then the first 100 levels were shifted to NIST values and retained in the rest of the calculation. The effective collision strengths for low-lying forbidden transitions are presented. In this paper, we are mostly interested in the rates at low temperatures, from 10 to 2,000 K, but we also include high-temperature results up to 100,000 K to compare with the plethora of previous calculations. It turns out that our results yield good agreement with some large-scale calculations even at high temperatures. 

We found the inclusion of $3d^64p$ is essential for reliable fine-structure transition data. In our 3-even BPRM/ICFT calculations when $3d^64p$ was not included, the dominant fine-structure transition $^6D_{9/2}-^6D_{7/2}$ was underestimated compared to other calculations, which is similar to the findings in \citet{bautista2015}. For the excitation from the ground level to the first excited level, most of their ICFT calculations give $\Upsilon$ (10$^4$~ K) of about 2, while their DARC calculation gives $\Upsilon$ (10$^4$~K) about 5. However, the electron configurations in their ICFT target are all of even parity, while their DARC target contained $3d^64p$. When $3d^64p$ was taken into consideration, our 3-mix/4-mix BPRM calculations are in good agreement with DARC calculations. It is an indication that the discrepancy in the work of \citet{bautista2015} likely depends on the difference in configuration expansion.  

The resulting level energies as well as Einstein A coefficients from the atomic structure calculations were used to evaluate the reliability of the target model. The GRASP$^0$, 3-mix and 4-mix AUTOSTRUCTURE target models could give good overall atomic structure. The effective collision strengths from the DARC calculations were adopted as the recommended values. The uncertainties were evaluated by calculating the standard deviation of 3-mix and 4-mix BPRM results from the recommended values. The complete data set is available online\footnote{Effective collision strengths in the Cloudy STOUT format can be obtained at www.physast.uga.edu/amdbs/excitation/.} in favor of astrophysical environment modeling.

\section*{Acknowledgements}
We would like to thank Dr. Connor Ballance for his assistance with the Dirac and BP R-matrix codes and calculations and Dr. Manuel Bautista for assistance with the AUTOSTRUCTURE package. BMMcL thanks the University of Georgia for the award of an adjunct professorship, and Queen's University for a visiting research fellowship.  Computing resources were provided by the Georgia Advanced Computing Resource Center, the UNLV National Super Computing Institute, and the UGA Center for Simulational Physics. This work was funded by NASA grant NNX15AE47G.

%%%%%%%%%%%%%%%%%%%%%%%%%%%%%%%%%%%%%%%%%%%%%%%%%%



\bibliographystyle{mnras}
\bibliography{bibtex} % if your bibtex file is called example.bib


%%%%%%%%%%%%%%%%%%%%%%%%%%%%%%%%%%%%%%%%%%%%%%%%%%
% Don't change these lines
\bsp	% typesetting comment
\label{lastpage}
\end{document}